\pdfoutput=1
\PassOptionsToPackage{pdftex,
pdfencoding=auto,
pdfnewwindow=true,
pdfusetitle=true,
bookmarks=true,
bookmarksnumbered=true,
bookmarksopen=false,
pdfpagemode=UseThumbs,
bookmarksopenlevel=1,
pdfpagelabels=true,
breaklinks=true,
colorlinks=true,
unicode=true,
pdfborder={0 0 0},
backref=false
}{hyperref}
\PassOptionsToPackage{usenames,dvipsnames,pdftex}{xcolor}
\documentclass[superscriptaddress,aps,pra,nofootinbib,onecolumn,twoside,reprint,letterpaper,longbibliography,showpacs,10pt,floatfix]{revtex4-2}
\usepackage[utf8]{inputenx}
\usepackage{amsmath,bbm}
\usepackage{amssymb}
\usepackage{graphicx}
\usepackage{color}
\usepackage{amsthm}
\usepackage{dsfont}
\usepackage{float}
\usepackage{cancel}
\usepackage{booktabs}
\usepackage[caption=false]{subfig}
\usepackage[alwaysadjust,flushleft,pointlessenum]{paralist}
\setdefaultenum{\bfseries1.}{}{}{}

\makeatletter 
    
\renewcommand\onecolumngrid{%
\do@columngrid{one}{\@ne}%
\def\set@footnotewidth{\onecolumngrid}%
\def\footnoterule{\kern-6pt\hrule width 1.5in\kern6pt}%
}

\renewcommand\twocolumngrid{%
        \def\footnoterule{%
        \dimen@\skip\footins\divide\dimen@\thr@@
        \kern-\dimen@\hrule width.5in\kern\dimen@}
        \do@columngrid{mlt}{\tw@}
}%

\makeatother

\usepackage[colorlinks=true, allcolors=blue]{hyperref}
\usepackage{amsthm}
\usepackage{amsmath}
\usepackage{amssymb}
\usepackage{graphicx}
\usepackage{color}
\usepackage{enumitem}

\newcommand{\W}{\mathrm{\W}}
\newcommand{\GHZ}{\mathrm{\GHZ}}

\newcommand{\beq}{\begin{equation}}
\newcommand{\eeq}{\end{equation}}
\newcommand{\beqa}{\begin{eqnarray}}
\newcommand{\eeqa}{\end{eqnarray}}

\newcommand{\ket}[1]{\ensuremath{\left|#1\right\rangle}}
\newcommand{\ketbra}[2]{\ensuremath{\left|#1\right\rangle\left\langle#2\right|}}

\newcommand{\norm}[1]{\left\lVert#1\right\rVert}

\renewcommand{\today}{\number\day\space\ifcase\month\or
   January\or February\or March\or April\or May\or June\or
   July\or August\or September\or October\or November\or December\fi
   \space\number\year}

\definecolor{myurlcolor}{rgb}{0,0,0.7}
\definecolor{myrefcolor}{rgb}{0.8,0,0}
\definecolor{purple}{RGB}{128,0,128}
\definecolor{ultramarine}{RGB}{63, 0, 255}
\definecolor{medblue}{RGB}{0, 0, 100}
\definecolor{googleblue}{RGB}{34, 0, 204}
\definecolor{panblue}{RGB}{0,24,150}
\definecolor{carmine}{RGB}{150, 0, 24}
\definecolor{gray}{RGB}{150, 150, 150}

\usepackage{amsmath,amsfonts,amssymb,amsthm,mathtools,bm,graphicx,verbatim,mathrsfs,units}
\usepackage{cprotect} %

\newtheoremstyle{defblock}{0.7\topsep}{0pt}{}{}{}{: }{0pt plus 1pt minus 1pt}{\thmname{\bfseries{#1}}\thmnumber{\bfseries{#2}}\color{medblue}\bfseries\thmnote{#3}}
\theoremstyle{defblock}

\theoremstyle{remark}

\newcommand{\same}[0]{{\textsf{Same}}}
\newcommand{\Bell}[0]{{\textsf{Bell}}}

\newcommand{\tagprop}[1]{\tag{\hyperref[#1]{P\ref{#1}}}}

\usepackage{microtype}
\microtypesetup{
expansion={true,nocompatibility},
protrusion={true,nocompatibility},
activate={true,nocompatibility},
tracking=true,
kerning=true,
spacing={true}
}
\usepackage[all=normal,floats=tight,mathspacing=tight,wordspacing=tight,paragraphs=normal,tracking=tight,charwidths=tight,mathdisplays=normal,sections=normal,margins=normal]{savetrees}
\everypar=\expandafter{\the\everypar\loosness=-1 }
\usepackage[style=american,autopunct=true]{csquotes}

\usepackage[scr=boondoxupr]{mathalfa} %

\usepackage{hyperref}
\hypersetup{
linkcolor=carmine,
citecolor=googleblue,
urlcolor=panblue,
anchorcolor=OliveGreen}

\begin{document}

\begin{abstract}
Quantum theory predicts the existence of genuinely tripartite-entangled states, %
which cannot be obtained from local operations over any bipartite entangled states and unlimited shared randomness.
Some of us recently proved that this feature is a fundamental signature of quantum theory. 
The state $\ket{{\rm GHZ}_3}=(\ket{000}+\ket{111})/\sqrt{2}$ gives rise to tripartite quantum correlations which cannot be explained by any causal theory limited to bipartite nonclassical common causes \emph{of any kind} (generalising entanglement) assisted with unlimited shared randomness.
Hence, any conceivable physical theory which would reproduce quantum predictions will necessarily include genuinely tripartite resources. 

In this work, we verify that such tripartite correlations are experimentally achievable.
We derive a new device-independent witness capable of falsifying causal theories wherein nonclassical resources are merely bipartite. Using a high-performance photonic $\ket{{\rm GHZ}_3}$ states with fidelities of $ 0.9741\pm0.002$, we provide a clear experimental violation of that witness by more than 26.3 standard deviation, under the locality and fair sampling assumption.
We generalise our work to the $\ket{{\rm GHZ}_4}$ state, obtaining correlations which cannot be explained by any causal theory limited to tripartite nonclassical common causes assisted with unlimited shared randomness.

\end{abstract}

\title{Experimental Demonstration that\\No Tripartite-Nonlocal Causal Theory Explains Nature's Correlations}

\date{\today}

\author{\underline{Huan Cao}}
\affiliation{CAS Key Laboratory of Quantum Information, University of Science and Technology of China, Hefei, 230026, China.}
\affiliation{CAS Center For Excellence in Quantum Information and Quantum Physics, University of Science and Technology of China, Hefei, 230026, China.}

\author{\underline{Marc-Olivier Renou}}
\email{marc-olivier.renou@icfo.eu}
\affiliation{ICFO-Institut de Ciencies Fotoniques, The Barcelona Institute of Science and Technology, Castelldefels (Barcelona), Spain.}

\author{\underline{Chao Zhang}}
\affiliation{CAS Key Laboratory of Quantum Information, University of Science and Technology of China, Hefei, 230026, China.}
\affiliation{CAS Center For Excellence in Quantum Information and Quantum Physics, University of Science and Technology of China, Hefei, 230026, China.}

\author{Gaël Massé}
\email{gael.masse@icfo.eu}
\affiliation{ICFO-Institut de Ciencies Fotoniques, The Barcelona Institute of Science and Technology, Castelldefels (Barcelona), Spain.}

\author{Xavier Coiteux-Roy}
\email{xavier.coiteux.roy@usi.ch}
\affiliation{Faculty of Informatics, Università della Svizzera italiana, Lugano, Switzerland.}

\author{Bi-Heng Liu} 
\affiliation{CAS Key Laboratory of Quantum Information, University of Science and Technology of China, Hefei, 230026, China.}
\affiliation{CAS Center For Excellence in Quantum Information and Quantum Physics, University of Science and Technology of China, Hefei, 230026, China.}

\author{Yun-Feng Huang}
\email{hyf@ustc.edu.cn}
\affiliation{CAS Key Laboratory of Quantum Information, University of Science and Technology of China, Hefei, 230026, China.}
\affiliation{CAS Center For Excellence in Quantum Information and Quantum Physics, University of Science and Technology of China, Hefei, 230026, China.}

\author{Chuan-Feng Li}
\email{cfli@ustc.edu.cn}
\affiliation{CAS Key Laboratory of Quantum Information, University of Science and Technology of China, Hefei, 230026, China.}
\affiliation{CAS Center For Excellence in Quantum Information and Quantum Physics, University of Science and Technology of China, Hefei, 230026, China.}

\author{Guang-Can Guo} 
\affiliation{CAS Key Laboratory of Quantum Information, University of Science and Technology of China, Hefei, 230026, China.}
\affiliation{CAS Center For Excellence in Quantum Information and Quantum Physics, University of Science and Technology of China, Hefei, 230026, China.}

\author{Elie Wolfe}
\email{ewolfe@perimeterinstitute.ca}
\affiliation{Perimeter Institute for Theoretical Physics, Waterloo, Ontario, Canada.}

\maketitle

\twocolumngrid

\emph{Introduction.---}
Since its inception, quantum theory has been remarkably successful in explaining and predicting various phenomena, such as the black-body radiation, the energy levels of the hydrogen atom, and more recently the violation of Bell inequalities~\cite{bell1964einstein}.
Every prediction of quantum theory which has been subjected to experimental testing has been validated, sometimes to record-breaking precision. General consensus among scientists is that the operational predictions of quantum theory are no longer in question, at least in regimes accessible with current technologies.

However, empirical success need not imply ontological truth. Accordingly, physicists remain motivated to explore alternatives to quantum theory, to either subsume or replace it.
There are several motivations for such considerations. 
Firstly, the mathematical formalism of quantum theory is the first representation of Nature to reject the intuitive local realistic approach of all previous physical theories. 
Secondly, efforts to construct 
alternatives to quantum theory have historically proven quite fruitful in enlightening our understanding of quantum theory by contrast \cite{SpekkensToyModel, Popescu1994}.
Finally, there are serious obstacles to reconciling quantum theory with general relativity \cite{Page2005}. Some paradoxes admit multiple potential resolutions, but there are ongoing efforts to either quantize gravity~\cite{Carlip2001,Hardy2019quantumEQ} or to generalize quantum theory~\cite{Smith2019,Galley2021NoGo}.

\begin{figure}
 
 \includegraphics[width=1\columnwidth]{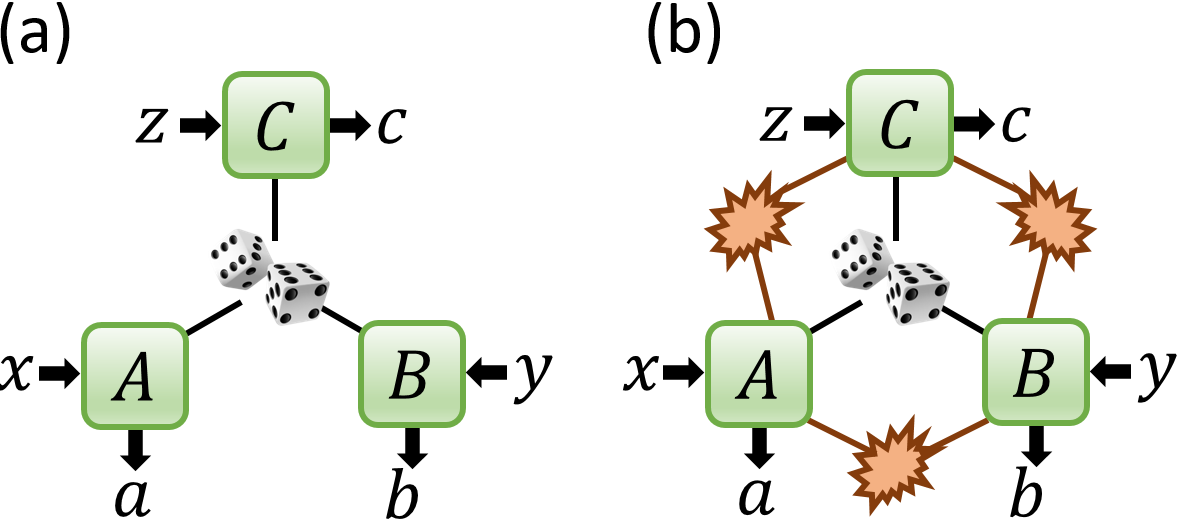}   
\caption{Bell’s theorem rules out any explanation of tripartite quantum correlations in terms of unbounded shared randomness (also called local-hidden-variable models), represented in (a) as dice. %
\hspace{0.5\textwidth}
In this letter we rule out the stronger model depicted in (b). Measuring a $\ket{\rm {\rm GHZ}_3}$ state, we obtain correlations that cannot be explained by any theory based on unbounded shared randomness (represented as dice) and any exotic generalisation of bipartite-entangled states (represented as starbursts).
We further generalise our setup to the $\ket{{\rm GHZ}_4}$ state, ruling out any theory based on unbounded shared randomness and generalisations of tripartite-entangled states. 
}
\label{fig:LHVModelOurModel}
\end{figure}

At the heart of what makes quantum theory nonclassical lies entanglement, a fundamental concept in quantum theory which underlies many non-classical behaviours (i.e. many quantum phenomena), such as nonlocality~\cite{Brunner2014}. The existence of this property of quantum correlations was first demonstrated by Bell with his eponymous theorem~\cite{bell1966lhvm,CHSHOriginal}. 
It shows that two parties measuring a maximally entangled source can obtain nonlocal correlations, which resist any explanation in terms of local-hidden-variable models. This was verified in several celebrated Bell experimental tests, namely in loophole-free CHSH violations~\cite{Freedman1972,Aspect1981,Tittel1998,Hensen2015,Shalm2015,Giustina2015,Rosenfeld2017,MingHan2018}. 

As a consequence, no predictive theory of Nature’s correlations can avoid introducing bipartite \emph{nonlocal} resources, namely states shared by two parties that admit no description in terms of classical randomness, \textit{i.e} that are not reproducible by local-hidden-variable models.
In the formalism of quantum theory this role is fulfilled by bipartite entanglement, but any potential theory that would replace quantum theory will necessary include a resource akin to quantum bipartite entanglement, such as the Popescu-Rohrlich box~\cite{Popescu1994}. 

However, the formalism of quantum theory admits nonclassical resources beyond bipartite entanglement. 
It postulates the existence of states which are so-called \emph{genuinely network tripartite-entangled}, i.e. states which cannot be obtained by three parties through Local Operations (LO) on bipartite entangled sources and Shared Randomness (SR)~\cite{NetworkEntanglement2020}. The $\ket{{\rm GHZ}_3}=\frac{1}{\sqrt{2}}(\ket{000}+\ket{111})$ state is such an example.
In other terms, quantum theory's mathematical formalism includes 3-way entanglement as an additional resource, strictly stronger than bipartite entanglement.
This greater resource underlies operational predictions that are qualitatively stronger than bipartite resources, such as a stronger version of Bell's theorem~\cite{Greenberger1990GHZ}.

Note that the intrinsicness of genuinely tripartite-entangled resources in quantum theory does not \emph{a priori} necessitate the existence of such resources in future alternatives to quantum theory. For instance, although $\ket{{\rm GHZ}_3}$ is required to achieve maximal violation of the Mermin inequality \emph{within quantum theory}, one can also reproduce the same operational correlations by means exclusively bipartite resources in alternative physical theories such as Boxworld~\cite{Barrett2005SvetFromPR,Contreras2021} (see Fig.~\ref{fig:LHVModelOurModel}b).
Might a future theory explain \emph{all} of Nature’s tripartite correlations without invoking any intrinsically tripartite nonclassical states? Formally, would it be possible to simulate all tripartite quantum correlations from theories restricted to the sort of bipartite exotic nonclassical states which generalising bipartite entanglement, supplemented with unlimited shared randomness?

In Refs.~\cite{Coiteux2021PRL,Coiteux2021PRA}, some of us developed a test to disprove any such explanation.
This test can be seen as a generalisation of Bell’s theorem: not only do we disprove any explanation in terms of unbounded shared randomness (also called local-hidden-variable models), but we also disprove any explanation relying on a yet-to-be-discovered causal theory that would involve exotic bipartite nonclassical resources and unbounded shared randomness, such as presented in Fig.~\ref{fig:LHVModelOurModel}b. 
Our test is predicated on the nonfanout inflation technique~\cite{Wolfe2016inflation}, which minimally assumes that a future theory should satisfy \emph{causality} and allow for \emph{device replication}~\cite{GisinNSI,Bancal2021Networks,Beigi2021,Pironio2021InPreparation}.

In this letter, our first main result is the computer-aided derivation of an improved test which is maximally robust against general noise models. The potential experimental demands for violating this new inequality are significantly reduced compared to the inequality derived in Ref.~\cite{Coiteux2021PRL}.
Then, in our second main result, we report an experimental demonstration of said inequality's violation using a photonic setup to experimentally exhibit correlations that cannot be explained by any theory based on shared randomness and bipartite nonlocal resources, however exotic. 
Furthermore, we expand both our witness derivation and experimental demonstration to the $\ket{{\rm GHZ}_4}=\frac{1}{\sqrt{2}}(\ket{0000}+\ket{1111})$ state, proving that it realizes correlations which cannot be explained neither by any theory relying on shared randomness and, this time, tripartite exotic nonlocal resources. Hence, we show experimentally that no tripartite-nonlocal causal theory can explain Nature's correlations.

\emph{The ideal protocol.---}
Here, we first propose a concrete experimental procedure which is a variant of the protocols of Refs.~\cite{Coiteux2021PRL,Coiteux2021PRA}. 
Then, we show that it is a valid test to disprove any explanation based on bipartite exotic nonclassical states and unlimited shared randomness. 

A $\ket{{\rm GHZ}_3}$ state is distributed to Alice, Bob, and Charlie.
The players perform independent measurements depending on their respective uniformly distributed random inputs $x,y,z\in\{0,1\}$. They produce respective outputs $a,b,c \in \{\pm 1\}$. Alice's and Charlie's respective measurements are $(\sigma_z,\sigma_x)$, and  Bob's respective measurements are $(\frac{\sigma_x+\sigma_z}{\sqrt{2}},\frac{\sigma_x-\sigma_z}{\sqrt{2}})$. 

In~Appendix~\ref{app:noise} we extend Algorithm~1 from Ref.~\cite[Sec.~IV]{Coiteux2021PRA} in order to obtain inequalities maximally robust against general noise models. %
We obtain the following novel inequality satisfied by any strategy based on any bipartite-nonlocal exotic resources and unrestricted shared randomness:
\begin{eqnarray}\label{eq:InequalityBipartiteExoticresources}\begin{aligned}
W_3\coloneqq 
&\langle A_0 B_0 \rangle + \langle B_0 C_0 \rangle  - \langle A_0 B_1 \rangle - \langle B_1 C_0 \rangle \\
& + 4 \langle A_0 C_0 \rangle+ 2\langle A_1 B_0 C_1 \rangle + 2 \langle A_1 B_1 C_1 \rangle 
~~\leq~ 8
\end{aligned}\end{eqnarray}
The considered tripartite quantum strategy gives a violation of ${W_3\to 4+4\sqrt{2}\approx 9.657}$. 
Under white noise $W_3$ evaluates to zero, implying a visibility against white noise of $v\gtrsim 0.828$.

 We generalize this to four players, with the shared quantum state $\ket{{\rm GHZ}_4}$, adding a fourth player named Dave. The generalized quantum strategy keeps Alice, Bob and Charlie as before, while Dave behaves as Charlie, measuring $(\sigma_z,\sigma_x)$ with inputs $w\in\{0,1\}$ and outputs $d\in\{\pm 1\}$. 
 Following the same technique, we obtain the following novel inequality satisfied by any strategy based on any tripartite-nonlocal exotic resources and unrestricted shared randomness:
 \begin{eqnarray}\label{eq:InequalityTripartiteExoticresources}\begin{aligned}
 W_4\coloneqq 
 \langle A_0 B_0 \rangle &- \langle A_ 0 B_1 \rangle + 2 \langle A_0 D_0 \rangle+2 \langle C_0 D_0 \rangle \\
&\hspace{1ex}  + \langle A_1 B_0 C_1 D_1 \rangle +\langle A_1 B_1 C_1 D_1 \rangle~~ \leq~ 6~~
\end{aligned}\end{eqnarray}
The quadpartite quantum strategy gives a violation of ${W_4\to 4+2\sqrt{2}\approx 6.828}$, with white noise visibility $v\gtrsim 0.879$.

\begin{figure}
    \centering
    \includegraphics[scale=0.08]{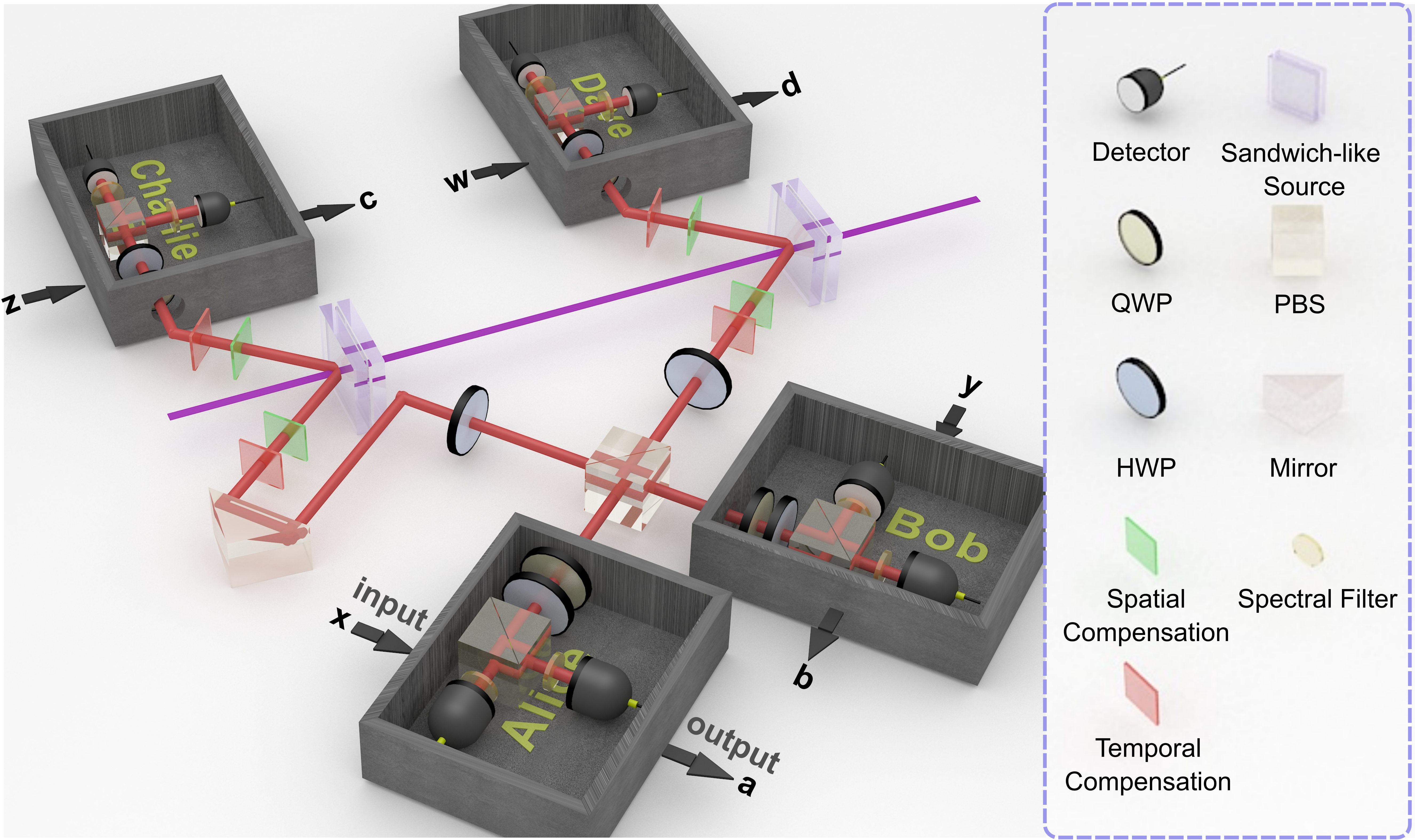}
    \caption{Scheme of the $\ket{{\rm GHZ}_4}$ experiment.
    The sandwich-like BBO resource consists of BBO-HWP-BBO configuration. In each source, two possible ways of down conversion may occur in former and latter BBO. They are made indistinguishable by the temporal and spatial compensation crystal in both arms. A 1nm (3nm) width spectral filter is inserted in the extraordinary (ordinary) photon before measurement to enhance the fidelity of the entangled state. The SPDC sources combined with the HOM interference corresponds to the multipartite entangled resources with shared randomness while the local measurement setting embedded in blacked boxes plays the role of the parties.
    For the $\ket{{\rm GHZ}_3}$ experiment, Dave projects over the diagonal basis $\ket{+}$.
    }
    \label{fig:setup}
\end{figure}

\emph{The experiment.---}
We experimentally implement this protocol in a photonic platform, based on the spontaneous parametric down-conversion process (SPDC). 
The experimental setup is illustrated in Fig.~\ref{fig:setup}. 
We first generate polarization twin photons by adopting the sandwich-geometry beam-like type-II SPDC \cite{GHZExperiment6Photons,wang2016experimental}. 
A frequency doubled pulsed ultraviolet laser with 390nm central wavelength is focused on sandwich-like combination of $\beta$-barium-borate (BBO) crystal to generate the maximally entangled state in polarization. 
A half-wave-plate (HWP) transforms the down-converted entangled photon into $\left|\psi\right\rangle =\left(\left|00\right\rangle+\left|11\right\rangle\right)/\sqrt{2}$. Here we encode logic qubit $\left|0\right\rangle$ ($\left|1\right\rangle$) in horizontal (vertical) polarization respectively.
For preparing a 4-photon (or 3-photon) GHZ state, two pairs of entangled states are generated by sequentially pumping two SPDC sources, and the extraordinary 
photons in each source are spatially overlapped in polarization beam splitter (PBS) to complete the Hong-Ou-Mandel (HOM) interference.
A trombone-arm delay line is introduced to make interfering photons indistinguishable in arrival time. The PBS acts as a parity check operation on interfering photons. The $\ket{{\rm GHZ}_4}$ state is obtained by collectively post-selecting on one photon for all parties.
Each party is capable of performing arbitrary local measurement on received qubit by polarization analyzing system consisting of a quarter-wave plate (QWP), a HWP, and a PBS. 
At last, the $\ket{{\rm GHZ}_3}$ state is obtained by post selecting on one of the photons being measured into diagonal basis $\left|+\right\rangle$.

\begin{figure}
    \centering
    \includegraphics[scale=0.125]{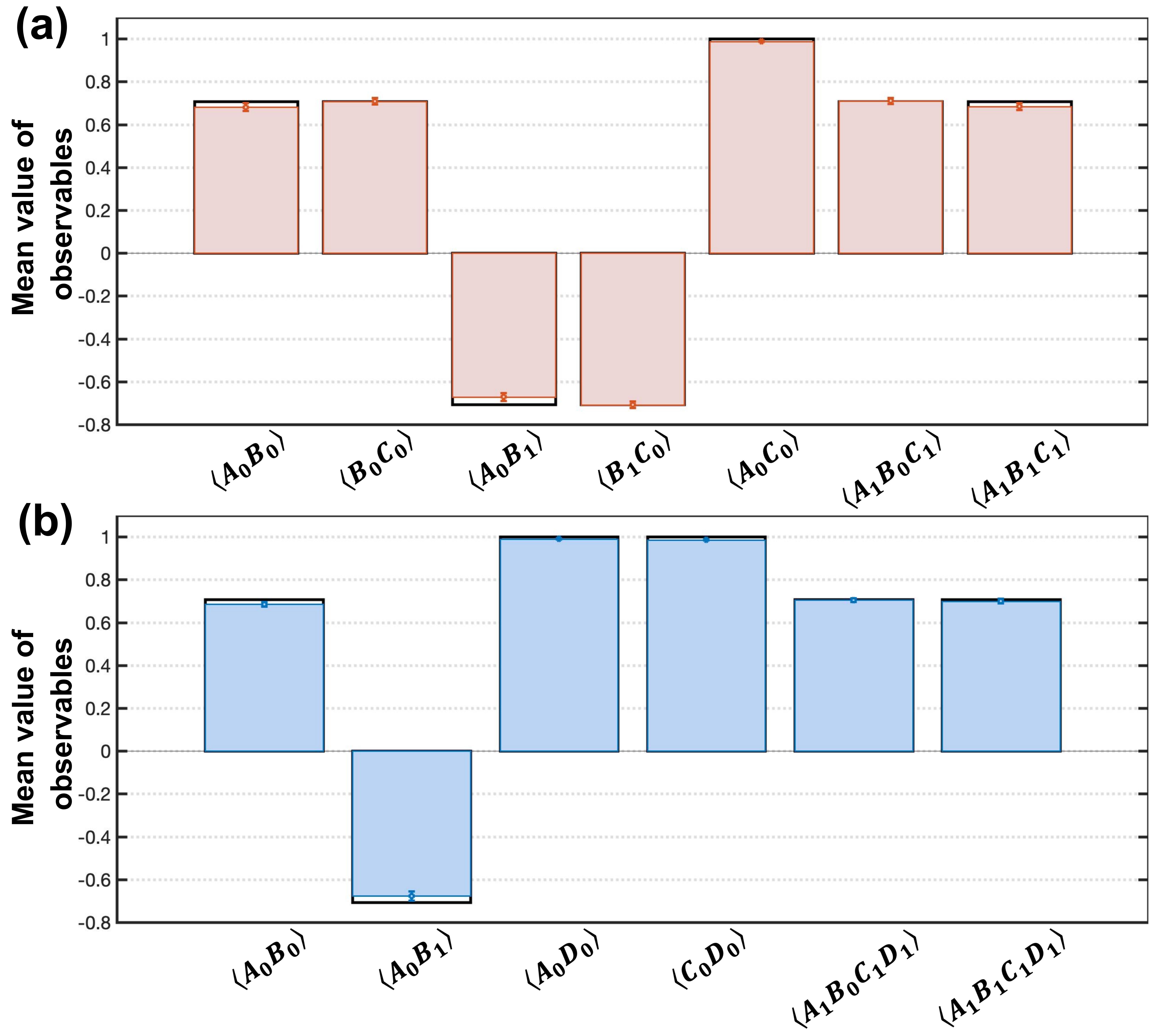}
    \caption{Experimental results for the 3-photon (a) and 4-photon (b) GHZ states, for all terms involved in the inequalities~\eqref{eq:InequalityBipartiteExoticresources} and~\eqref{eq:InequalityTripartiteExoticresources} . 
    Theoretically predictions are plotted with transparent pillars and experimental values in nontransparent ones. The errorbars are computed through a Monte Carlo simulation to evaluate the Poisson noise of the photonic statistics.}
    \label{fig:ExperimentResult}
\end{figure}

\emph{The results.---}
In designing an experiment to achieved the requisite fidelity, we note that the dominant obstacles to inequality violation is white noise related to higher-order emission. To suppress the higher-order emission noise we therefore maintain the experimental operation in low pumping power yielding an approximate four-fold counting rate of 1 click/s. The quantum measurement strategy was optimized to maximally violate the inequality, following the aforementioned protocol. 

We evaluate the experimental violations of the two inequalities given by Eq.~\eqref{eq:InequalityBipartiteExoticresources} and Eq.~\eqref{eq:InequalityTripartiteExoticresources} for our experimental sources of 3-photon and 4-photon GHZ states, terms by terms. The mean values of each term is given in Fig.~\ref{fig:ExperimentResult}.
Thanks to the high performance of our sources, our experimental mean values are in good agreement with the theoretical predictions for  ideal GHZ states. 
We obtain the violations
\begin{align}
W_3=9.5150\pm 0.0576,
\end{align}
with more than 26.3 standard deviation, and
\begin{align}
W_4=6.7154\pm 0.0256,
\end{align}
with more than 27.9 standard deviation.
These significant violations provide strong support for the idea that correlations achieved via 3-photon (and 4-photon) GHZ state cannot be explained by bipartite-nonlocal (resp. tripartite-nonlocal) causal theories.

\emph{Discussion.---}
First, note that our protocol can be generalised to any number of parties. 
For instance, one could in principle generalise our experiment to the $\ket{{\rm GHZ}_5}$ state. Of course, to do so we would first need to isolate a suitable inequality to test against. We found our  algorithmic approach to inequality derivation to be intractable for such large $N$, even with symmetry reduction. We have identified a new family of inequalities which is slightly too demanding in terms of noise with respect to our experimental setup (see discussion in~Appendix~\ref{app:another_ineq}).
We suspect that it is possible to obtain new inequalities, maybe including extra measurements for the parties, which would reduce the critical fidelity requirements to what is achievable with current technologies for the $\ket{{\rm GHZ}_5}$ and $\ket{{\rm GHZ}_6}$ states.

Our experimental demonstration admits two common loopholes considered in Bell nonlocality experiments: the locality loophole and the postselection loophole. We now briefly discuss the theoretical reasons why these loopholes might undermine our conclusions, why we consider these loopholes implausible, and how these loopholes might be closed in future experiments.

In an alternate universe in which Nature would not admit genuinely multipartite nonlocal sources, our experiment can still be simulated if any party's measurement outcomes were allowed to causally depend on the other parties' choices of measurement settings.
This is our locality loophole.  
It is standard to \emph{assume} locality in a Bell experiment. 
Ghostly influence from one measurement apparatus to another is theoretically conceivable, but practically implausible. 
Nevertheless, the locality loophole can in-principle be closed in a future experiment, by enforcing a space-like separation of the parties, which is accessible to current technologies \cite{erven2014experimental}.

Another loophole which could explain away our experimental data without requiring genuinely multipartite nonlocal sources is the postselection loophole. After all, in our final analysis, we exclusively considered the subset of trials in which all every party received exactly a photon. Postselection could artificially simulate our experimental data if the detection versus nondetection determination for a given party is allowed to causally depend on that party's choice of measurement setting.
The assumption that each party's `detection' random variable is causally independent of their setting variable is known as \emph{fair sampling}; the fair sampling assumption is quite commonly employed in photonic experimental investigations of nonlocality.
However, fair sampling \emph{by itself} does not entirely close the postselection loophole. Indeed, one can simulate all statistics arising from measurements on $\ket{{\rm GHZ}_3}$ using \emph{strictly bipartite sources} if one allows for postselection, even under the constraint that the postselection variable is causally independent of the measurement settings.\footnote{Let $A$, $B$ and $C$ share three bipartite maximally entangled states $\ket{\phi^+}_{A_1B_2}, \ket{\phi^+}_{B_1C_2},\ket{\phi^+}_{C_1A_2}$ where $\ket{\phi^+}=\frac{1}{\sqrt{2}}(\ket{00}+\ket{11})$. Let each party $P\in\{A,B,C\}$ non-destructively measure the parity operators $\Pi_{\rm even}^{P_1P_2}=\ketbra{00}{00}_{P_1P_2}+\ketbra{11}{11}_{P_1P_2}, \Pi_{\rm odd}^{P_1P_2}=\ketbra{01}{01}_{P_1P_2}+\ketbra{10}{10}_{P_1P_2}$ and postselect on the $\Pi_{\rm even}^{P_1P_2}$ event. The (cumulatively) postselected state is $\frac{1}{\sqrt{2}}(\ket{00}_{A}\ket{00}_{B}\ket{00}_{C}+\ket{11}_{A}\ket{11}_{B}\ket{11}_{C})$, which is equivalent to  $\ket{{\rm GHZ}_3}$. Note that we have enforced causal independence of postselection from measurement settings, since in this protocol each party may obtain their individual postselection status prior to specifying their measurement setting.}
As such, the conclusion that our experimental results demonstrate the presence of genuinely multipartite nonlocal sources hinges on one additional assumption, beyond just locality and fair sampling. This further assumption is that it is possible --- in the exotic foil theories under consideration --- to move any postselection process \emph{from the measurements to the sources} without modifying the measurement statistics. This theory-agnostic assumption codifies our belief that --- even though \emph{we} utilized postselection at the measurements --- Nature would allow for a reproduction of our experiment with postselection at the sources, i.e., with \emph{heralded event-ready sources}.\footnote{We illustrate this argument in quantum theory, keeping the same example as before. 
It is \emph{in principle} possible in quantum theory (albeit technologically challenging) to move all the three $\ket{\phi^+}$ sources and the nondestructive parity measurements to a single common location without modifying the experimental statistics. 
This would effectively convert the previous example into an event-ready protocol, in which a successful state preparation is heralded by the $\Pi_{\rm even}^{P_1P_2}$ events. Notably, one can leverage event-ready sources to close the postselection loophole by ensuring that the choice of input for each party is decided after the heralding.}

This postselection loophole might be closed in a future experiment by preparing the $\ket{{\rm GHZ}_3}$ (resp. $\ket{{\rm GHZ}_4}$) state in a heralded event-ready preparation using three (resp. four) cascaded SPDC sources. 
This would generalise the work of~\cite{Hamel2014GHZ3MerminSvetlichny}, where two cascaded SPDC sources are used to obtain a heralded EPR pair. Current hardware, however, is insufficient for preparing heralded multipartite entanglement sources capable of achieving required noise tolerance and creation rates.

\emph{Conclusion.---}
In this letter, we experimentally demonstrated that any causal theory 
that aims to explain all possible  correlations of Nature must include genuinely tripartite-nonlocal and four-partite-nonlocal states. 
In quantum theory, this role is taken by what we call genuinely LOSR (network) tripartite and fourpartite entangled states already introduced in Ref.~\cite{NetworkEntanglement2020}. 
We have also shown how to improve the algorithm of~\cite{Coiteux2021PRA} to obtain inequalities maximally robust against general noise models and white noise models, reducing the demand in computing power by exploiting the symmetries of the problem.
We expect this method to be fundamental for future (loophole free) demonstrations of Nature multipartie nonlocal genuiness. 

Note that this LOSR-based definition is more restrictive than the traditional LOCC-based definition of genuine tripartite entanglement due to Seevinck and Uffink~\cite{Seevinck2001}. 
Let us first emphasize that this traditional definition is not adapted to the no-signalling context.
Indeed, a four-qubit state composed of a singlet shared between Alice and Bob as well as a singlet shared between Bob and Charlie satisfies Seevinck's criterion for genuinely tripartite entanglement, despite being created from bipartite entanglement\footnote{Note that to mask the bipartite entanglement in the state's structure, Bob could in principle unitarily transfer his state to a four level atomic system}. 
In the same way, the definition of LOSR multipartite genuine nonlocality that we adopt in our letter is more restrictive than the traditional definition of genuine tripartite nonlocality due to Svetlichny~\cite{Svetlichny}. The traditional definition is susceptible to precisely the same sort of hacking: the tripartite correlations obtained from two parallel CHSH violations between Alice and Bob as well as between Bob and Charlie are genuinely tripartite nonlocal according to Svetlichny's criterion.

These two historical definitions of genuine multipartiteness for nonlocality and entanglement are ill-suited for our study because they were motivated by quantifying resourcefulness relative to Local Operations and
Classical Communication (LOCC). However, when analysing Bell-inequality violations, we assume that the involved parties are spacelike separated, and this enforces the No-Signalling condition. When classical communication is
forbidden, the only form of nonclassical-resources processing that remains is via Local Operations and Shared Randomness (LOSR). 
This new approach is closely related to the concept of network nonlocality which
has been extensively studied in the past decade~\cite{TavakoliReview,fritz2012bell,Branciard2010,Renou2019}.

The use of the traditional notions of multipartite genuine entanglement and nonlocality for analysing the states and correlations obtained in various multipartite experiments is widespread. 
Let us conclude by saying that the use of these notions should be questioned in light of the context of the experiment in which they are used, keeping in mind that they are not adapted to the No-signalling context~\footnote{In particular,~\cite{Contreras2021} `demonstrates' that multipartite genuine nonlocality can be obtained from bipartite pure entangled states distributed in a network, considering the historical Svetlichny definition for multipartite genuine nonlocality. This paradoxical statement (bipartite entanglement can be used to create multipartite genuine nonlocality) is solved if one consider our more appropriate (in the considered context) redefinition of multipartite genuine entanglement.}~\cite{Barrett2005,Gallego2012}.

\emph{Note added.---} While finishing this manuscript, we became aware of related work by Ya-Li Mao et al.~\cite{mao2022test}.

\emph{Acknowledgements.---}
We thank Antonio Acín and Enky Oudot for valuable discussions.
This research was supported by the Swiss National Science Foundation (SNF) and Perimeter Institute for Theoretical Physics. Research at Perimeter Institute is supported in part by the Government of Canada through the Department of Innovation, Science and Economic Development Canada and by the Province of Ontario through the Ministry of Colleges and Universities. 
M.-O.R. is supported by the Swiss National Fund Early Mobility Grants P2GEP2\_19144 and the grant PCI2021-122022-2B financed by MCIN/AEI/10.13039/501100011033 and by the European Union NextGenerationEU/PRTR, and acknowledges the Government of Spain (FIS2020-TRANQI and Severo Ochoa CEX2019-000910-S [MCIN/ AEI/10.13039/501100011033]), Fundació Cellex, Fundació Mir-Puig, Generalitat de Catalunya (CERCA, AGAUR SGR 1381) and the ERC AdG CERQUTE.
This research was supported by the National Key Research and Development Program of China (No. 2017YFA0304100), 
the National Natural Science Foundation of China (Nos. 11774335, 11821404, 11734015, 62075208).

\emph{Contributions.---}
H.C., M-O.R. and C.Z contributed equally to this work.

\bigskip
\nocite{apsrev42Control}
\setlength{\bibsep}{1pt plus 1pt minus 1pt}
\bibliographystyle{apsrev4-2-wolfe}
\bibliography{biblio}

\appendix
\onecolumngrid

\clearpage
\section{Experimental details}\label{app:expdetails}

{\it EPR source.---} A sandwich-like structure crystal is used to generate polarization-entangled photon pairs, which consists of two adjacent, 1-mm-thick $\beta$-barium borate (BBO) crystals and a true-zero-order half-wave plate (THWP) inserted in the middle. The two BBO crystals (BBO1 and BBO2) are identically cut with beamlike type-II phase-matching and their optical axes are parallel in the horizontal plane. Thus BBO1 and BBO2 both produce photon pairs in the polarization state $|H\rangle_e|V\rangle_o$, the subscript o (e) denotes the ordinary (extraordinary) photon with respect to the crystal. The THWP rotates the polarization of the photon pairs produced by BBO1 to the orthogonal state $|V\rangle_e|H\rangle_o$. When the two possible ways of generating photon pairs (in BBO1 and BBO2 respectively) are made indistinguishable by spatial and temporal compensations, the photon pairs are prepared in entangled state $\frac{1}{\sqrt{2}}(|H\rangle_e|V\rangle_o-|V\rangle_e|H\rangle_o)$.

{\it GHZ states preparation.---} As shown in Fig.1 of the main text, we use two EPR sources to produce the three- and four-photon GHZ states. The laser pulse from a mode-locked Ti:sapphire laser (with a central wavelength of 780 nm, a pulse duration of 140 fs, a repetition rate of 80 MHz) is first frequency doubled to 390 nm and then pumps the two sources sequentially. We initialize the two EPR sources to both produce polarization state $\frac{1}{\sqrt{2}}(|H\rangle_e|H\rangle_o+|V\rangle_e|V\rangle_o)$. Then the two e-polarized photons are directed to overlap on a PBS. It is possible to check that when there is one and only one photon in each output, only two terms $|HHHH\rangle$ and $|VVVV\rangle$ are post-selected. The two terms are further made indistinguishable by interference on the PBS, thus the initial product state is projected into the four-photon GHZ state $\frac{1}{\sqrt{2}}(|HHHH\rangle+|VVVV\rangle)$. The three-photon GHZ state can be prepared by projecting the last photon on $|+\rangle$. In our experiment, the postselection is realized by post-processing the detected data after the experiment is finished. Note that by using auxiliary photons and quantum teleportation we can detect a single photon without destroying it and keeping its quantum information intact, thus the postselection can be realized in principle in the state preparation part and lead to a heralded GHZ state.

{\it Detection.---} In the measurement part, each photon is first passed through an interference filter (IF), we use 1- and 3-nm bandwidths filters for e- and o-polarized photons respectively. Then each photon is detected by a polarization analysis system which consists of one QWP, one HWP, one PBS and two fiber-coupled single-photon detectors. In the experiment, we use a low pump power of 60 mW to limit the higher-order emission noise in SPDC. The two-photon counting rate is about 12000 Hz and four-photon counting is about 1 Hz. To measure the LOSR inequalities, we perform several local measurement settings, and  record all combinations of the four-fold coincidence events by using a coincidence unit (UQDevice), with a coincidence window of 4 ns. The raw data is shown in Table I.

\begin{table}[htb]
\centering
\caption{Raw data for local measurement settings. X and Z are Pauli operators. Here we use ``+(-)" to denote the +1(-1) port detector firing for each photon. The data collection time for ZZZZ and ZZZX is 20000s, the other settings is 4000s. }
\begin{tabular}{|c|c|c|c|c|c|c|c|c|c|c|c|c|c|c|c|c|} \hline

Setting &++++&+++-&++-+&++- -&+-++&+-+-&+- -+&+- - -&-+++&-++-&-+-+&-+- -&- -++&- -+-&- - -+&- - - - \\\hline
$X\frac{(Z+X)}{\sqrt{2}}XX$     &439    &72    &84   &414    &67   &516   &412    &69    &94   &374   &402    &74   &433    &83    &43   &389 \\\hline
$X\frac{(Z-X)}{\sqrt{2}}XX$    &62   &372   &376    &73   &400    &77    &71   &413   &356    &59    &79   &371    &67   &454   &354    &65 \\\hline
$Z\frac{(Z+X)}{\sqrt{2}}XX$    &390   &371   &395   &336    &78    &78    &81    &66    &53    &69    &71    &56   &364   &372   &351   &371 \\\hline
$Z\frac{(Z-X)}{\sqrt{2}}XX$   &369   &366   &411   &384    &87    &70    &75    &97    &67    &59    &56    &53   &351   &367   &337   &359 \\\hline
$XXXX$   &450     &5    &11   &451     &9   &513   &531     &7     &4   &430   &428     &9   &499     &5     &9   &461 \\\hline
$ZZZZ$  &8552          &16          &13           &0           &9          &11          &14          &18          &15         & 19         & 11 &13           &0          &19          &20        &8311\\\hline
$ZZZX$   &5229        &4651           &7           &6          &19          &22&20          &24          &19          &21          &28          &20&11          &10        &4426        &4861\\\hline
$X\frac{(Z+X)}{\sqrt{2}}ZX$   &403        &406           &72           &83          &73          &71 &412          &427          &443          &429          &74          &66  &68          &81        &438        &422\\\hline
$X\frac{(Z-X)}{\sqrt{2}}ZX$   &422        &443          &85           &66          &66          &62   &424          &400          &415          &420          &76          &82   &62          &71        &394        &422\\\hline
\end{tabular}
\end{table}

\begin{figure}
    \centering
    \includegraphics[scale=0.35]{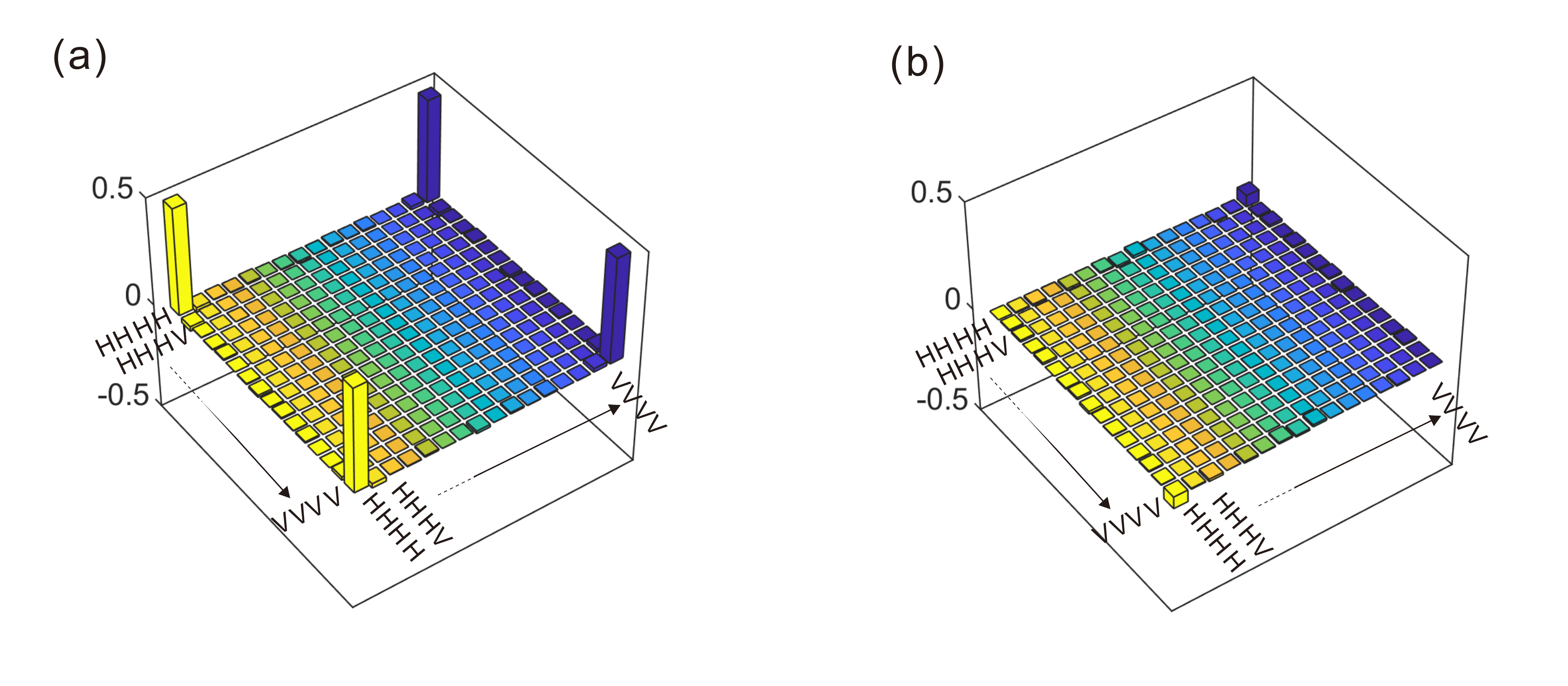}
    \caption{(a) Real and (b) Imaginary part of the reconstructed density matrix of the prepared four-photon GHZ state.}
    \label{fig:tomoGHZ4}
\end{figure}

{\it Characterization of the prepared states.---} From the measurement result of XXXX, we know that the HOM interference visibility in our experiment is at least $0.9691\pm0.0041$. From the measurement results of XXXX and ZZZZ, we can give a lower bound on the state fidelity according to the two setting entanglement witness
\begin{equation}
    \mathcal{W}_{\mathrm{GHZ}_{\mathrm{N}}}:=3 \mathbb{I}-2\left[\frac{S_1^{\left(\mathrm{GHZ}_{N}\right)}+\mathbb{I}}{2}+\prod_{k=2}^{N} \frac{S_k^{\left(\mathrm{GHZ}_{N}\right)}+\mathbb{I}}{2}\right]
\end{equation}
where
\begin{equation}
\begin{array}{l}
S_1^{\left(\mathrm{GHZ}_{\mathrm{N}}\right)}:=\prod_{k=1}^{N} X^{(k)} \\
S_k^{\left(\mathrm{GHZ}_{\mathrm{N}}\right)}:=Z^{(k-1)} Z^{(k)} \quad \text { for } k=2, 3,\ldots,N.
\end{array}
\end{equation}
are stabilizers for GHZ state. The measured result is $\langle \mathcal{W}_{\mathrm{GHZ}_{\mathrm{4}}} \rangle=-0.9482\pm0.0048$. Then we can deduce the state fidelity $F=\langle \mathrm{GHZ}_4|\rho_{\mathrm{exp}}|\mathrm{GHZ}_4\rangle>(1-\langle \mathcal{W}_{\mathrm{GHZ}_{\mathrm{N}}} \rangle)/2=0.9741\pm0.0024$.
We also perform a quantum state tomography to characterize the prepared state, the reconstructed density matrix is shown in Fig.\ref{fig:tomoGHZ4}. The state fidelity is calculated to be $0.9762\pm0.0032$, which agrees with the above result very well.

\section{General Robustness and Specific Noise Models}\label{app:noise}

  This discussion presumes prior understanding of Algorithm 1 from Ref.~\cite[Sec.~IV]{Coiteux2021PRA}.
There, some of us introduced a technique to find constraints over distributions obtainable from $(N{-}1)$-way nonlocal resources and unrestricted shared randomness.
 Given a nonfanout inflation of the $(N{-}1)$-way nonlocal resources model (see Ref.~\cite[Fig.~1]{Coiteux2021PRA}), we exhibit a Linear Programming (LP) test providing necessary \emph{satisfiability} constraints. 
More precisely, if a given correlation $P_{A_1...A_N|X_1...X_N}$ can be obtained from $(N{-}1)$-way nonlocal resources and unrestricted shared randomness, one could in principle duplicate the building blocks used to obtain it and replace them in the nonfanout inflated scenario.
In this new scenario, one would observe a new correlation satisfying some linear compatibility constraint with the inflated scenario itself, and with $P_{A_1...A_N|X_1...X_N}$.
If these linear constraints cannot be satisfied, then it proves that $P_{A_1...A_N|X_1...X_N}$ is not compatible with $(N{-}1)$-way nonlocal resources and unrestricted shared randomness.
These linear constraint form the following LP:

\begin{samepage}
\medskip
\noindent$\textbf{LP}_{sat}$: 
\par\nopagebreak We say that $\vec{P}\propto \textbf{LP}_{sat}$ if and only if
\begin{compactitem}
\item there exists some non-negative vector $\vec{x}$ such that
\item for matrix $M_1$ we satisfy $M_1\cdot \vec{x}=\vec{P}$ and 
\item for matrix $M_2$ we satisfy $M_2\cdot \vec{x}=\vec{0}$.
\end{compactitem}
\noindent When referring to \emph{normalized} mutlipartite correlation vectors, we use the notation ${\vec{P}\in \textbf{LP}_{sat}}$ instead of ${\vec{P}\propto \textbf{LP}_{sat}}$.\footnote{A mutlipartite correlation vector is a family of conditional probability distributions. We use the terminology ``$\vec{P}$ is normalized" and the notation ${\norm*{\vec{P}}{=}1}$ to indicate that $\vec{P}$ has total unit probability along every conditional specification. Note that in $\textbf{LP}_{sat}$ the normalization of $\vec{x}$ implicitly depends on the normalization of $\vec{P}$.}
\medskip
\end{samepage}

In this notation $M_1$ is a marginalization matrix capturating the fact that the observed probabilities must arise as the marginals of some nonfanout inflation, and $M_2$ is a matrix capturing other equality constraints such as the fact that all inflations admit nonsignalling correlations, and that various marginals of different nonfanout inflations must coincide; see  Ref.~\cite[Algorithm~1]{Coiteux2021PRA}.
Of course, the linear program implied by nonfanout inflation is itself a relaxation of true causal compatibility. That is, $\vec{P}\not\in \textsf{LOSR\_Genuinely\_Multipartite}\implies \vec{P}\in\textbf{LP}_{sat}$, but not not vice versa.

In this work, however, we do not merely want to test whether or not a given correlation vector $\vec{P}$ satisfies the linear programming constraints. Instead, our goal is to use linear programming to extract an inequality showing that $\vec{P}$ \emph{cannot} satisfy the linear programming constraints, and moreover our inequality should be maximally robust against general noise.

Firstly, let us recast the satisfiability problem as an optimization problem. The visibility $v^\star(\vec{P})$ against general noise models is the largest real number $v$ such that ${v{\times}\vec{P}{+}(1{-}v){\times} \vec{P'}} \not\in \textsf{LOSR\_Genuinely\_Multipartite}$ (for \emph{any} normalized correlation $\vec{P'}$).\footnote{\label{footnote:monotones} The \textbf{resource robustness} of $P$ is equal to one minus the visibility of $P$ against general noise. Resource robustness is a measure of how distinguishable a non-free object is from a set of free objects in any convex-linear resource theory. In our case, the object is a mutlipartite correlation vector, and the free set are all correlations which are \emph{not} genuinely LOSR multipartite. Resource robustness shows in standard Bell nonlocality, where the free set is the set of Bell local correlation. It also appears in entanglement theory, as a measure of distinguishability of an entangled state from the set of separable states. Ref.~\cite[Sec.~4.2.5]{gonda2021monotones} does highlights how this sort of monotone is generic in convex-linear resource theories, see especially Fig.~4b there. 
\par Another paradigmatic monotone in convex-linear resource theories is \textbf{resource weight}. In Bell nonlocality, for instance, this measure is known as the \emph{nonlocal fraction}. To study genuine LOSR multipartiteness with such a measure we simply adjust the choice of free objects as appropriate. That is, we can define a measure called \emph{genuine LOSR multipartite fraction} ($\textbf{GMF}_\text{LOSR}$) such that ${\textbf{GMF}_\text{LOSR}(\vec{P})\coloneqq  \displaystyle\min_{w,\vec{P'}\vec{P''}} \left(w\;\middle|\; \vec{P}=w{\times}\vec{P'}+(1{-}w){\times}\vec{P''},\, \vec{P''}\not\in \textsf{LOSR\_Genuinely\_Multipartite}\right)}$. See Ref.~\cite[Fig.~4a]{gonda2021monotones} for visualization.} 
Relative to the LP relaxation, however, we define  $v^\star_\text{LP}(\vec{P})$ as the largest real number $v$ such that ${v{\times}\vec{P}{+}(1{-}v){\times} \vec{P'}} \in \textbf{LP}_{sat}$.
Linearity of the definition implies that one can restrict $\vec{P'}$ to be normalized and nonsignalling whenever $\vec{P}$ is normalized and nonsignalling without loss of generality, since all correlations in the set $\textbf{LP}_{sat}$ are normalized and nonsignalling.  Plainly $v^\star_\text{LP}(\vec{P})\geq v^\star(\vec{P})$. However, even this inflation-related optimization problem is apparently nonlinear, as it involves a product of the target variable $v$ with the unspecified probabilities in $\vec{P'}$. We get around this by recognizing that the following formulations of $v^\star_\text{LP}(\vec{P})$ are equivalent.
\begin{subequations}\begin{align}
v^\star_\text{LP}(\vec{P})&\coloneqq \max_{v,\vec{P'}} \left(v\;\middle|\; \vec{P'}\geq \vec{0},\,{v{\times}\vec{P}{+}(1{-}v){\times} \vec{P'}} \in \textbf{LP}_{sat} \right),&&\text{ with }\norm*{\vec{P'}}{=}1\text{ implicit,}\\
&= \max_{v,\vec{s}} \left(v\;\middle|\;\vec{s}\geq\vec{0},\,{v{\times}\vec{P}{+}\vec{s}} \in \textbf{LP}_{sat} \right),&&\text{with }\norm*{\vec{s}}{=}1{-}v\text{ implicit,}\\
&= \max_{\vec{s}} \left(\norm*{\vec{P}{+}\vec{s}}^{-1}\;\middle|\;\vec{s}\geq\vec{0},\,{\vec{P}{+}\vec{s}} \propto \textbf{LP}_{sat} \right),&&\text{with unrestricted }\norm*{\vec{s}}{\in}[0,\inf)\,,\\
&= \left(\min_{\vec{s}} \left(\norm*{\vec{P}{+}\vec{s}}\;\middle|\;\vec{s}\geq\vec{0},\,{\vec{P}{+}\vec{s}} \propto \textbf{LP}_{sat} \right)\right)^{-1}
\end{align}\end{subequations}
We therefore implement the final formulation, making use of a constant-valued vector $\vec{c}$ such that $\vec{c}\cdot \vec{x}=1$ if $\vec{x}$ is normalized.\footnote{More generally, this formulation can be adapted to compute the resource robustness monotone via linear programming whenever both the enveloping set of object and the subset of free objects are polytopes. In our case, the set of free objects is \emph{outer approximated} by a polytope, and thus we obtain \emph{lower bounds} on the true resource robustness, i.e. upper bounds on the true visibility against general noise. In fact, the resource \emph{weight} monotone~\cite{gonda2021monotones} can also be cast as a linear program whenever resource theory is polytopic. As an example, the genuine LOSR multipartite fraction introduced in footnote~ref{footnote:monotones} can be lower bounded by simply flipping the inequality signs around in $\textbf{LP}_{opt, primal}$. That is, ${\textbf{GMF}_\text{LOSR}(\vec{P})\geq 1- \displaystyle\max_{\vec{x}} \left(\vec{c}\cdot\vec{x}\;\middle|\; \vec{x}\geq \vec{0},\, M_1\cdot \vec{x}\leq\vec{P},\, M_2\cdot \vec{x}=\vec{0}\right)}$.}

\begin{samepage}
\medskip
\noindent$\textbf{LP}_{opt, primal}$:%
\par\nopagebreak The visibility $v^\star_\text{LP}(\vec{P})$ against general noise models is equal to $\frac{1}{\tau_{\text{primal}}(\vec{P})}$, where $\tau_{\text{primal}}(\vec{P})\coloneqq{\displaystyle\min_{\vec{x}}\;\vec{c}\cdot \vec{x}}$ such that
\begin{compactitem}
\item $\vec{x}\geq \vec{0}$ and
\item $M_1\cdot \vec{x}\geq\vec{P}$ and 
\item $M_2\cdot \vec{x}=\vec{0}$.
\end{compactitem}
\medskip
\end{samepage}

We can now \emph{dualize} that linear program. Let $\vec{y_1}$ be the dual vector associated with the rows of $M_1$, and let  $\vec{y_2}$ be the dual vector associated with the rows of $M_2$. Then,

\begin{samepage}
\medskip
\noindent$\textbf{LP}_{opt, dual}$: 
 \par\nopagebreak %
The visibility $v^\star_\text{LP}(\vec{P})$ against general noise models is equal to $\frac{1}{\tau_{\text{dual}}(\vec{P})}$, where $\tau_{\text{dual}}(\vec{P})\coloneqq{\displaystyle\max_{\vec{y_1},\vec{y_2}}\; \vec{y_1}\cdot \vec{P}}$ such that
\begin{compactitem}
\item $\vec{y_1}\geq \vec{0}$ and $\vec{y_2}\geq \vec{0}$ and
\item $M_1^T \cdot \vec{y_1} + M_2^T \cdot \vec{y_2} \leq \vec{c}$.
\end{compactitem}
\medskip
\end{samepage}

LP duality guarantees that $\tau_{\text{primal}}(\vec{P})=\tau_{\text{dual}}(\vec{P})$ \cite{LPDuality}. The inequality resulting from the dual formulation is that $\vec{y_1}\cdot \vec{P}\leq 1$ for all $\vec{P}\in \textbf{LP}_{sat}$, and moreover, for $\vec{P}\not\in \textbf{LP}_{sat}$ we have $\vec{y_1}\cdot \vec{P}=\frac{1}{v^\star_\text{LP}(\vec{P})}>1$. Note that our inequalities must have nonnegative coefficients associated with event probabilities. The negative coefficients for the inequalities presented in the main text (as well as the bounds being different from unity) are consequences of the authors' choice to represent the inequalities in terms of expectation values instead of probabilities, motivated by compactness of presentation. 

The visibility of a correlations against \emph{general} noise models is generally  more demanding (i.e., higher visibility) than against \emph{particular} noise models. Visibility against a particular noise model $P'$ is given by 
\begin{align}\begin{split}
v^\star(\vec{P},\vec{P'})&\coloneqq \max_v \left(v\;\middle|\;{v{\times}\vec{P}{+}(1{-}v){\times} \vec{P'}} \not\in \textsf{LOSR\_Genuinely\_Multipartite}\right)\\
&\leq v^\star_\text{LP}(\vec{P},\vec{P'})\coloneqq \max_{v} \left(v \;\middle|\; {v{\times}\vec{P}{+}(1{-}v){\times} \vec{P'}} \in \textbf{LP}_{sat}\right).
\end{split}\end{align}
By contrast, visibility against \emph{general} noise is given by ${v^\star(\vec{P})=\max_{\vec{P'}} v^\star(\vec{P})\leq v^\star_\text{LP}(\vec{P})= \max_{\vec{P'}} v^\star_\text{LP}(\vec{P})}$.

Given a multipartitness witness of the form $\vec{y}\cdot \vec{P}\leq b$, it is straightforward to verify an upper bound on the general (or specific) visibility of any incompatible nonsignalling correlation. That is, 
\begin{align}\label{eq:specificnoise}
\text{ If } &\vec{y}\cdot \vec{P}\leq b \text{ for all } \vec{P}\in \textbf{LP}_{sat}\nonumber\\
\text{ then }&{v^\star(\vec{P},\vec{P'})\leq v^\star_\text{LP}(\vec{P},\vec{P'})\leq \max_v \left(v \;\middle|\; v\times\vec{y}\cdot \vec{P}+(1-v)\times\vec{P'}\leq b\right)},
\end{align}
and similarly
 ${v^\star(\vec{P})\leq \max_{\vec{P'}} v^\star(\vec{P},\vec{P'})}$.
 An inequality is \emph{optimal} relative to general or specific noise models if the visibility bound implied by the given inequality cannot be improved upon by any \emph{other} valid inequality.

In  Ref.~\cite{Coiteux2021PRA} some of us reported that there exists tripartite quantumly-realizable correlations with visibility \emph{against white noise} of $\approx 0.828$. 
Readers of this article can confirm that finding by examining $W_3$ here. 
Indeed, one can readily see that that $W_3$ evaluates to zero on white noise. 
Since $W_3$ evaluates to $4+4\sqrt{2}$ on our quantum correlations, it follows from Eq.~\eqref{eq:specificnoise} that $W_3$ implies a visibility against white noise of $v^\star\gtrsim 0.828$. Since the minimum evaluation of $W_3$ under all nonsignalling correlations is -8, we evidently have a visibility against worst-case noise models of $v^\star\gtrsim 0.906$, by solving ${v{\times}(4+4\sqrt{2})+(1{-}v){\times}(-8)=8}$.

\subsection{Optimally witnessing genuine LOSR multipartiteness with quanutum correlations}

The inequalities presented in the main text are \emph{optimal} relative to general noise. That is, they are explicitly derived using $\textbf{LP}_{opt, dual}$. In an experimental implementation, however, we are concerned about \emph{specific} noise models. We did not verify the optimality of these inequalities relative to our explicit noise models.

To discover the inequalities presented in the main text we used certain quantum correlations which were known to be genuinely multipartite LOSR nonlocal. That is, each inequality was discovered by seeding $\textbf{LP}_{opt, dual}$ with a particular quantumly-realizable $\vec{P}$. One might wonder if perhaps our discovered inequalities can be violated \emph{further} by \emph{different} quantum strategies. That is, our inequalities are certainly optimal for our seed quantum correlations, but perhaps a different quantum correlation is optimal for a given inequality.
To find the best inequality and quantum correlation \emph{pair} we have to essentially perform concurrent optimization of $\vec{y_{1}}$ (and $\vec{y_2}$) \emph{and} $\vec{P}$. That is: ${\displaystyle \max_{\vec{y_1},\vec{y_2},\vec{P}} \left(\vec{y_1}\cdot\vec{P}\;\middle|\; \vec{y_1}\geq \vec{0},\,\vec{y_2}\geq \vec{0},\, M_1^T \cdot \vec{y_1} + M_2^T \cdot \vec{y_2} \leq \vec{c},\,\vec{P}\in\textsf{Quantumly\_realizable}\right)}$. This could be approximated by a see-saw algorithm. Happily, we verified that our initial quantum correlations are stable under subsequent optimization. That is, we employed the analytic quantum bound optimization technique of Ref.~\cite{AnalyticQuantumBounds} to confirm that no quantum correlation can increase the violation of our derived inequalities beyond the values achieved by our initial quantum strategies. This gives us strong confidence that we have identified truly optimal quantum strategies to target for experimentally simulation, and that $W_3$ and $W_4$ are similarly truly optimal inequalities to test the experimental data against.

It is likely, however, that inequalities exhibiting even better noise resistance could be found if one allows more inputs and/or outputs for the parties.

\section{Noise robustness: Improvements of the new inequality}\label{app:fidelity}
Ref.~\cite{Coiteux2021PRA} reported that genuine LOSR tripartite nonlocality could in-principle be witnessed using fidelity with the 3-photon GHZ state of $85\%$. That visibility calculation, however, is based on assuming a noise model of
\begin{equation}\label{eq:whitenoise}
    \rho_{p}=p\left|\mathrm{GHZ_{N}}\right\rangle\left\langle \mathrm{GHZ_{N}}\right|+(1-p)\frac{\mathrm{I}_N}{2^{N}},
\end{equation}
corresponding to $p\approx 0.828$ and GHZ fidelity given by $f=p+\frac{1-p}{2^N}$.

In our experiment, however, a more accurate model is to take both dephasing noise and white noise into account. In such case the noisy GHZ state is modeled as
\begin{align}\begin{aligned}\label{eq:realnoise}
    \rho_{p,k}=p\times\left|\mathrm{GHZ_{N}}\right\rangle \left\langle \mathrm{GHZ_{N}}\right|+k(1-p)\times\left|\mathrm{GHZ_{N}^{-}}\right\rangle \left\langle \mathrm{GHZ_{N}^{-}}\right|+(1-p)(1-k)\times\frac{\mathrm{I}_N}{2^{N}}\,,
\end{aligned}\end{align}
where the phase flipped terms
\begin{align*}
\left|\mathrm{GHZ_{N}^{-}}\right\rangle =\frac{\left|0\right\rangle ^{\otimes N}-\left|1\right\rangle ^{\otimes N}}{\sqrt{2}}
\end{align*}
introduces the dephasing noise, which stems from imperfection of the Hong-Ou-Mandel interference, and the white noise is closely related to the higher-order emission noise. Fidelity with the $N$-partite GHZ state in this noise model is given by ${f=p+\frac{(1-p)(1-k)}{2^N}}$. 
The only inequality available in Ref.~\cite{Coiteux2021PRA} (Eq.~(1)) gives a threshold value of $p\approx 0.879$ for such states.

\begin{figure}
     \centering
     \includegraphics[scale=0.113]{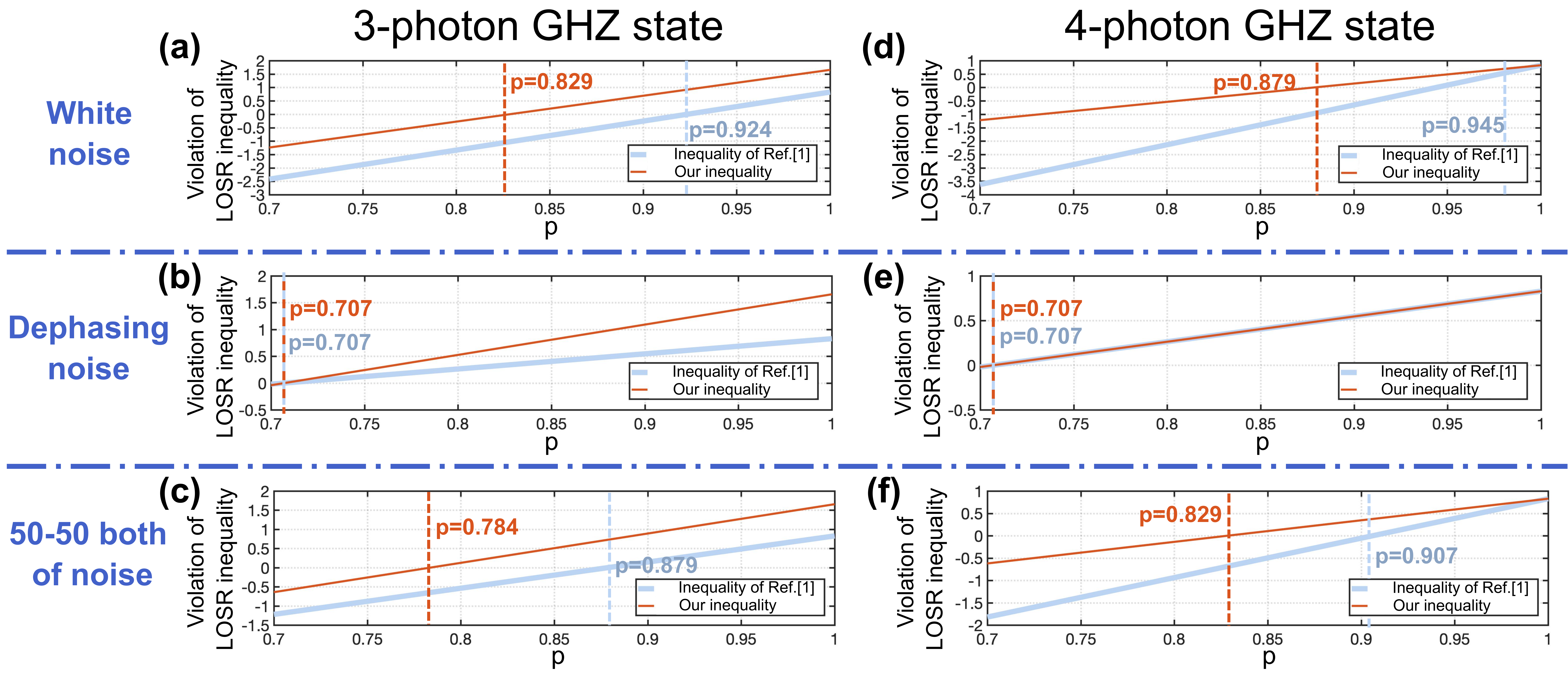}
     \caption{Comparison of the violation of the inequalities of inequalities Eq.~(1) and Eq.~(14) (for $N=4$) of Ref.~\cite{Coiteux2021PRA} (in blue) with our improved inequalities (1) and (2) reported in the main paper (in red), with noisy GHZ state in the case of the 3-photon GHZ state (a-c) and the 4-photon GHZ state (d-f). For plots (a), (d), we adopt a white noise model ($k=0$). For plots (b), (e), we adopt a dephasing noise model ($k=1$). For plots (c), (f), we adopt a mixed noise model ($k=0.5$). Red lines represent the eventual violation of the inequalities Eq.~(1) and Eq.~(2), the violation threshold is indicated by a red dotted line. Blue lines represent the eventual violation of the main text inequalities Eq.~(1) and Eq.~(14) (for $N=4$) of Ref.~\cite{Coiteux2021PRA}, the violation threshold is indicated by a blue dotted line. All inequalities are rescaled by a constant such that the violation is obtained for positive values.
}
     \label{fig:LOSRnoise}
 \end{figure}

We now explicitly compare the performence of the inequalities Eq.~(1) and Eq.~(14) (for $N=4$) of Ref.~\cite{Coiteux2021PRA} with our improved inequalities (1) and (2) reported in the main paper. In Fig.~\ref{fig:LOSRnoise}, we consider the violation threshold under different type of noise, for $p$ from 0.7 to 1: (i) GHZ state with pure white noise (i.e. taking $k=1$ in eq.\ref{eq:realnoise}), (ii) GHZ state with pure dephasing noise (i.e. taking $k=0$ in eq.\ref{eq:realnoise}), (iii) GHZ state with both white and dephasing noise (i.e. taking $k=0.5$ in eq.\ref{eq:realnoise}). This last choice is the closest to our experimental conditions.
The figure shows that our improved LOSR always exhibits a higher noise tolerance than the inequality of \cite{Coiteux2021PRA}, except in the (nonrealistic case) of $k=0$. In the case of $k=0.5$, our new inequality improves the threshold from $p=0.879$ to $p=0.784$ for 3-partite (corresponding to a fidelity threshold improved from $88.66\%$ to $79.75\%$), and $p=0.907$ to $p=0.829$ for 4-partite (corresponding to a fidelity threshold improved from $90.99\%$ to $83.43\%$).

Since, in practice, we can experimentally achieve much higher fidelities with our photonic platform, we proceeded with the implementation with high confidence.

\section{$N$-party inequality}\label{app:another_ineq}
We derive analytically an inequality for any $N$ parties that is robust to noise. We take the two games in Reference~\cite{Coiteux2021PRA} (we use the same notation),
\begin{align}\label{eq:BellStandard-N}
I_\Bell^{\tilde{C}_1{=}1}\coloneqq  &\langle A_0B_0 \rangle_{\tilde{C}_1{=}1} + \langle A_0B_1 \rangle_{\tilde{C}_1{=}1}+\langle A_1B_0 \rangle_{\tilde{C}_1{=}1} -\langle A_1B_1 \rangle_{\tilde{C}_1{=}1} \,,
\end{align}
and
\begin{align}\label{eq:defsame-N}
I_{\same_N}\coloneqq \langle A_0B_2\rangle +& \langle B_2C_{0[1]}\rangle+ \langle C_{0[1]}C_{0[2]}\rangle+[\dots]+ \langle C_{0[N-3]}C_{0[N-2]}\rangle\,.
\end{align}

We then take Eq.~15 from Reference~\cite{Coiteux2021PRA} (which reformulates
Theorem~1 of Ref.~\citep[Eq.~(11)]{AugusiakMonogamy}),
\begin{equation}\label{eq:monogamy-N}
I_{\Bell}^{\tilde{C}^1_1{=}1}\circ \{A^1B^1\} + 2\langle A_0^1C_{0[N-2]}^2\rangle_{\tilde{C}^1_1{=}1} \le  4 \,.
\end{equation}

We can obtain an inequality that is more robust to noise by observing that the Bell inequality still holds when conditioning the CHSH inequality upon the collectives of Charlies. \textit{i.e.}, if we define
\begin{align}\label{eq:BellStandard-N2}
I_\Bell^{\tilde{C}_1{=}-1}\coloneqq  &\langle A_0B_1 \rangle_{\tilde{C}_1{=}-1} + \langle A_0B_0 \rangle_{\tilde{C}_1{=}-1}+\langle A_1B_1 \rangle_{\tilde{C}_1{=}-1} -\langle A_1B_0 \rangle_{\tilde{C}_1{=}-1} \,,
\end{align}
then Eq.~\ref{eq:monogamy-N} implies that the following convex combination also holds,
\begin{equation}\label{eq:monogamy-N2}
\frac{1-\langle \tilde{C}_1\rangle}{2}I_{\Bell}^{\tilde{C}^1_1{=}-1}\circ \{A^1B^1\}+\frac{1+\langle \tilde{C}_1\rangle}{2}I_{\Bell}^{\tilde{C}^1_1{=}1}\circ \{A^1B^1\} + 2\langle A_0^1C_{0[N-2]}^2\rangle \le  4 \,.
\end{equation}

We combine Eq.~\ref{eq:monogamy-N} with the following equation (Eq.~20 from Reference~\cite{Coiteux2021PRA}),
\begin{equation}
\langle A^2_0C^2_{0[N-2]}\rangle\geq I_\same\circ \{A^2B^2C^2_{[0]}\dots C^2_{[N-2]}\} - N+2\,,\label{eq:transitivity-A-N}
\end{equation}
to obtain the following noise-robust inequality
\begin{equation}\label{eq:monogamy-N3}
\frac{1-\langle \tilde{C}_1\rangle}{2}I_{\Bell}^{\tilde{C}^1_1{=}-1}\circ \{A^1B^1\}+\frac{1+\langle \tilde{C}_1\rangle}{2}I_{\Bell}^{\tilde{C}^1_1{=}1}\circ \{A^1B^1\} + 2I_\same\circ \{A^2B^2C^2_{[0]}\dots C^2_{[N-2]}\} \le  2N \,.
\end{equation}

This inequality is violated up to $2\sqrt{2}+2(N-1)$ with the $\ket{\textrm{GHZ}_N}$ states when Alice and Bob do, respectively, the measurements $\{\sigma_Z,\sigma_X\}$ and $\{\frac{\sigma_X+\sigma_Z}{\sqrt{2}},\frac{-\sigma_X+\sigma_Z}{\sqrt{2}},\sigma_Z\}$, and all other players (i.e. the Charlies) measure following $\{\sigma_Z,\sigma_X\}$.

Using the white-noise model of Eq.~\ref{eq:whitenoise},
we obtain a violation down to $p>\frac{n}{n-1+\sqrt{2}}$,
while using the dephasing-plus-white-noise model of Eq.~\ref{eq:realnoise} results in a violation for the regime $p>\frac{k+2N(1-k)}{2\sqrt{2}-2(1+kN)+k+2N}$.

For $N=5$, this implies noise robustness of $p>\frac{5}{4+\sqrt{2}}\approx 0.923$ for the white noise model, which correspond to minimal fidelity requirements of $f\approx 0.926$.
Considering the experimentally more realistic case of $k=0.5$, we obtain a noise robustness of $p>\frac{5.5}{3.5+2\sqrt{2}}\approx 0.869$, which correspond to minimal fidelity requirements of $f\approx 0.873$.
This fidelity bound is on the edge of what can be achieved with current state-of-the-art multi-photon technologies \cite{Chao2019demonstrating,li2020photonic}, which makes the observation of a violation challenging. Moreover, this would require very low counting rates to keep preserve the high fidelity of the source.

\end{document}